\documentclass[%
 reprint,
superscriptaddress,
 amsmath,amssymb,
 aps,
prb,
]{revtex4-2}

\usepackage{graphicx}
\usepackage{dcolumn}
\usepackage{bm}
\usepackage{hyperref}
\usepackage{color}
\usepackage{subfigure}
\begin{document}

\preprint{APS/123-QED}

\title{Thermal Stoner-Wohlfarth Model for Magnetodynamics of Single Domain Nanoparticles: Implementation and Validation}

\author{Deniz Mostarac}%
\affiliation{%
Physics Department, University of Roma “La Sapienza”, Rome, Italy \\ Computational and Soft Matter Physics, University of Vienna, Vienna, Austria}
\email{deniz.mostarac@univie.ac.at}
\author{Andrey A. Kuznetsov}
\affiliation{%
 Computational and Soft Matter Physics, University of Vienna, Vienna, Austria}%
\date{\today}
\author{Santiago Helbig}
\affiliation{%
 Physics of Functional Materials, University of Vienna, Vienna, Austria}%
\author{Claas Abert}
\affiliation{%
 Physics of Functional Materials, University of Vienna, Vienna, Austria}%
\author{Pedro A. S\'anchez}
\affiliation{%
Computational and Soft Matter Physics, University of Vienna, Vienna, Austria}%
\author{Dieter Suess}
\affiliation{%
 Physics of Functional Materials, University of Vienna, Vienna, Austria}%
\author{Sofia S. Kantorovich}
\affiliation{%
 Computational and Soft Matter Physics, University of Vienna, Vienna, Austria}%
\date{\today}

\begin{abstract}
We present the thermal Stoner-Wohlfarth (tSW) model and apply it in the context of Molecular Dynamics simulations. The model is validated against an ensemble of immobilized, randomly oriented uniaxial particles (solid superparamagnet) and a classical dilute ferrofluid for different combinations of anisotropy strength and magnetic field/moment coupling, at a fixed temperature. We compare analytical and simulation results to quantify the viability of the tSW model in reproducing the equilibrium (with and without dipole-dipole interactions) and dynamic (without dipole-dipole interactions) properties of magnetic soft matter systems. We show that if the anisotropy of a particle is more than five times higher than the thermal fluctuations, tSW is applicable and efficient. This approach allows one to consider the interplay between Néel and Brownian relaxation, often neglected in the fixed point-dipole representation-based magnetic soft matter theoretical investigations.
\end{abstract}

\maketitle

\section{Introduction}\label{sec:intro}
The idea to control materials using magnetic fields, inspired in the early 20th century by the works of Langevin~\cite{langevin1905a,langevin1905b} and through pioneering works on magnetic fluids by Elmore~\cite{elmore38a} and later by Resler and Rosenzweig~\cite{resler64a}, has since manifested a branch of science studying magnetic soft matter. The fundamental insight from these pioneering works is that small ferromagnetic particles suspended in a magneto-passive liquid represented a paramagnetic system whose magnetization can be described with a Langevin function. In other words, one understood that it is possible to engineer and appropriate the magnetic response of complex systems from the collective magnetic response of constituent nanoparticles that themselves have well defined and finely controllable magnetic properties. Under the umbrella term of magnetic soft matter, various magnetoresponsive systems have been explored, including ferrofluids,\cite{resler1964magnetocaloric,cebers1997magnetic,odenbach2008ferrofluids,2009-odenbach} ferrogels,\cite{zrinyi1998kinetics,weeber2018polymer} elastomers,\cite{volkova2017motion,filipcsei2007magnetic,li2014state,odenbach2016microstructure,sanchez2019surface} magnetic gels~\cite{frank1993voltage,zrinyi1998kinetics,Weeber_2012} and magnetic filaments (MFs)~\cite{Dreyfus_2005,2008-benkoski}, with a research action focused on generating a macroscopic response by coupling the magnetic response of magnetic nanoparticles, that can be affected via magnetic fields in a contained and precise manner, mechanically or chemically to its environment, such as a polymeric matrix or a viscous environment in general. This class of problem lends itself particularly for classical simulation studies such as Molecular Dynamics (MD). 
The key that unlocked such simulation studies of  composite, complex magnetoresponsive systems, their magnetic properties and response to magnetic fields, is the fixed point dipole representation~\cite{gubin2009magnetic}. In this picture, the dipole moment is depicted as vector that is always coaligned with the anisotropy axis of the nanoparticle, with a fixed magnitude and orientation in the particle reference frame. The fixed point dipole representation of magnetic colloids in general is at the heart of the overwhelming majority of theoretical and classical simulation studies of magnetic fluids and magnetic soft matter~\cite{ivanov2023magnetization,mostarac2022rheology,mostarac2022nanopolymers,novikau2022behaviour,rosenberg2023influence}.
It is a powerful and flexible representation, that albeit undoubtedly practical, incurs important assumptions and approximations. 
There are no degrees of freedom for the dipole moment to exploit with respect to the anisotropy axis. In other words, magnetic relaxation can only manifest through physical rotation of the nanoparticle body, commonly refereed to as Brownian relaxation. Strictly speaking, the fixed point dipole representation is appropriate for single-domain nanoparticles with infinitely high uniaxial anisotropy~\cite{coffey2012thermal}. 
If one is only interested in applied field coupling and far field interactions, more complex magnetic colloids can also be represented as fixed point dipoles, as long as they can be considered as ideally ferro-/ferrimagnetic~\cite{buyevich1992equilibrium}. Clearly, the fixed point dipole representation fundamentally limits the scope of the studies based on it. In general, depending on the colloids size and/or material, it can neither be taken as given that the dipole moment is coaligned with the anisotropy axis, nor that it has a fixed orientation with respect to it. In fact, the dipole moment has degrees of freedom within the nanoparticle body, which manifest as additional magnetic relaxation mechanisms, most prominent of which is commonly refereed to as Néel relaxation. Generally, the dipole moment states within a nanoparticle correspond to local free energy minima. With Néel relaxation one implies that thermal fluctuations can cause jumps between the available states. In other words, the dipole moment can, driven by thermal fluctuations, flip its orientation with respect to the easy axis. In this context, the anisotropy energy can be interpreted as an energy barrier that needs to be surmounted in order for such a flip to occur. The Brownian and N\'eel relaxation are two major mechanisms in the description of the magnetic properties of single-domain nanoparticles, and are well captured with a combination of the ideas presented in the seminal works of~\citet{stoner1948mechanism}, \citet{neel1949influence} and \citet{brown1963}.
When modeling the dynamics of magnetic soft matter, capturing Brownian and N\'eel relaxation is fundamentally important~\cite{taukulis2012coupled,ilg2017equilibrium,bender2018a,elfimova2019static,ilg2022longest,poperechny2023multipeak}. Beyond the academic interest in accurately describing the dynamics of magnetic nanoparticles, understanding magnetic relaxation opens numerous avenues for technological applications~\cite{2004-menager-pol,2015-backes-jpcb,mandal19a,sung20a,cao20a,gao20a,biglione20a}. Among others, the interplay between different relaxation mechanisms is key for optimizing magnetic hyperthermia~\cite{hilger12iron,durr2013magnetic,perigo2015fundamentals,ABENOJAR2016440}.

In this work, we present the thermal Stoner-Wohlfarth (tSW) model as a low computational cost method that allows one to include explicit and accurate magnetodynamics into simulation studies of magnetic soft matter with single domain, uniaxial nanoparticles. In this context, magnetodynamics entails: (i) Brownian relaxation; (ii) Néel relaxation mechanisms, and (iii) the ability of a dipole moment not to be co-aligned with the nanoparticle anisotropy axis in the presence of magnetic and/or dipole fields. Talking about low computational cost methods to include explicit magnetodynamics, one must mention the highly efficient and accurate approaches to numerical solve of the Fokker-Planck equation~\cite{poperechny2020combined,kroger2022combined}, for the so-called “Egg model” of superparamagnetic particle dynamics initially developed by~\citet{shliomis1994theory}. While solving the Fokker-Planck equation is a very powerful and fast approach to solve magnetodynamics, it is restricted to relatively simple non-interacting systems, and in general, is less flexible than stochastic simulations. It is also worth highlighting the diffusion jump model~\cite{ilg2019diffusion,ilg2020dynamics}, as an example of a similar, low computational cost method to simulate magnetodynamics. This model however, does not capture the (iii) aspect of magnetodynamics present in the tSW model, because in it, the magnetic moment is necessarily aligned with the crystallographic axis of the colloid.

The paper is structured as follows: We first introduce a formal problem description in Section \ref{sec:intro}, with a detailed discussion on the relevant energy (\ref{ssec::energy}) and timescales (\ref{ssec::time}). In Section \ref{sec:results}, the model is validated against: (a) an ensemble of immobilized, randomly oriented uniaxial particles (solid superparamagnet) and (b) a classical dilute ferrofluid. This is equivalent to making the distinction on the basis of whether the nanoparticles exhibit only Néel relaxation mechanisms or both Brownian and Néel relaxation mechanisms, respectively. In both cases, we compare analytical and simulation results to quantify the viability of the thermal SW model in reproducing equilibrium (magnetization for a non-interacting and interacting system; \ref{ssec::equil}) and dynamic (susceptibility for a non-interacting system; \ref{ssec::susc}) properties of magnetic soft matter systems. Note that throughout the manuscript, unless explicitly stated otherwise, we are referring to non-interacting systems. A detailed account of the tSW model implementation, simulation details and units is presented in Section \ref{ssec:model}. The theoretical framework behind the tSW model has been introduced in \citet{chuev2007nanomagnetism}, where it is called the generalized SW model. The generalized SW model has been implemented and used in several studies under the name of kinetic Monte Carlo (kMC) approach.~\cite{chantrell2000calculations,tan2014magnetic,ruta2015unified,jonasson2019modelling,wolfschwenger2024molecular} However, the kMC terminology has been used inconsistently in the literature. It is important to clarify that, broadly speaking, the kMC approach implies that the magnetic degrees of freedom are governed by sampling the magnetic energy in the SW model via a MC algorithm. The transition probabilities are calculated in accordance with the generalized SW model. The kMC approach is, however, not limited to a two state approximation. In other words, in the kMC approach, the dipole moment does not need to point in accordance with the minima of the magnetic energy. Therefore, it should be considered as distinct from both the generalized SW and the tSW model. The tSW approach can be seen as a subset of kMC, in the sense that it strictly adheres to the two state approximation. However, the tSW model can also be seen as a generalization of the kMC approach, in the sense that couples it with Langevin Dynamics. Regardless, in this contribution, we present the first systematic study of the applicability of the tSW model with respect to anisotropy energy/temperature and magnetic field strength, where the viability and applicability of the model in the context of MD simulations of magnetic soft matter is qualified.

\subsection{Formal problem description}\label{ssec::energy}

\begin{figure}[!h]
     \centering
     \includegraphics[width=0.5\linewidth]{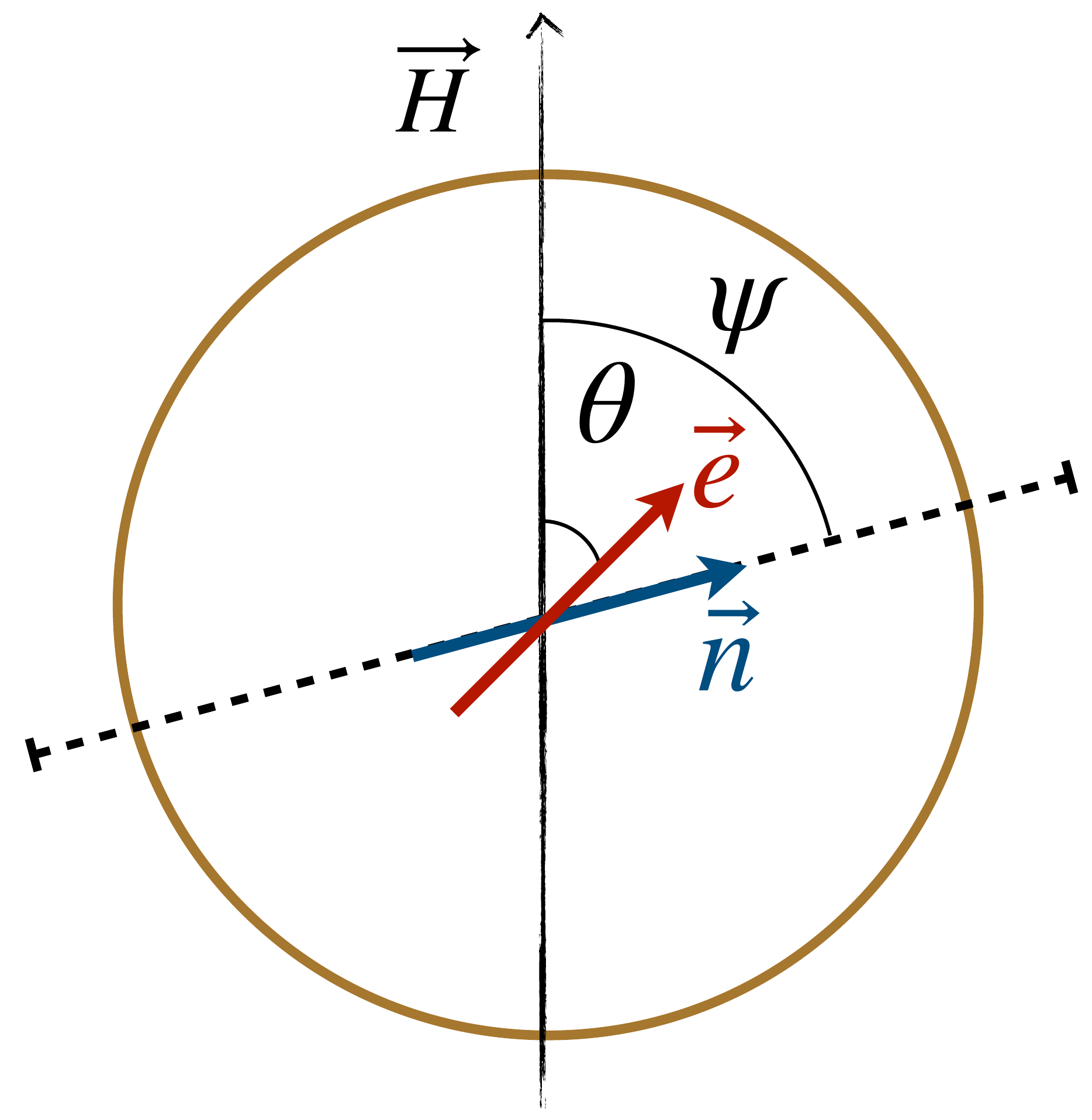}
     \caption{Geometric details including angles used in the formulation of the total magnetic energy  in the Stoner-Wohlfarth model.}
     \label{fig:doodle}
 \end{figure}

A magnetic nanoparticle is subjected to a uniform magnetic field $\vec{H}$ at a fixed temperature $T$. The classical, \emph{athermal} formulation of the total magnetic energy of a nanoparticle was introduced in~\citet{stoner1948mechanism}
\begin{equation}\label{eq:sw_mag_ene}
    U =  - \mu_0 \mu(\vec{e}\cdot\vec{H}) - KV (\vec{e} \cdot \vec{n})^2,
\end{equation}
where:
 \begin{itemize}\setlength\itemsep{0.2em}
    \item $\mu_0$ is the  vacuum magnetic permeability;
    \item $\vec{e} = \vec{\mu}/ \mu = (\cos \phi \sin \theta, \sin \phi \sin \theta, \cos \theta)$ is the unit vector of the magnetic moment $\vec{\mu}$, where $\mu = |\vec{\mu}| =  M_s V$, $\theta$ is the angle between the magnetic moment and the field and $\phi$ is the corresponding polar angle, for notations see Fig.~\ref{fig:doodle};
    \item $M_s$ is the saturation magnetization of the particle material;
    \item $V = (\pi/6)d^3$ is the particle volume, $d$ is the particle diameter;
    \item $K$ is the particle anisotropy constant;
    \item $\vec{n} = (\cos \zeta \sin \psi, \sin \zeta \sin \psi, \cos \psi)$ is the unit vector of the anisotropy axis, where $\psi$ is the angle between the applied field and the anisotropy axis and $\zeta$ is the corresponding polar angle, see Fig.~\ref{fig:doodle}.
    
 \end{itemize} 

\begin{figure}
 \includegraphics[width=0.93\linewidth]{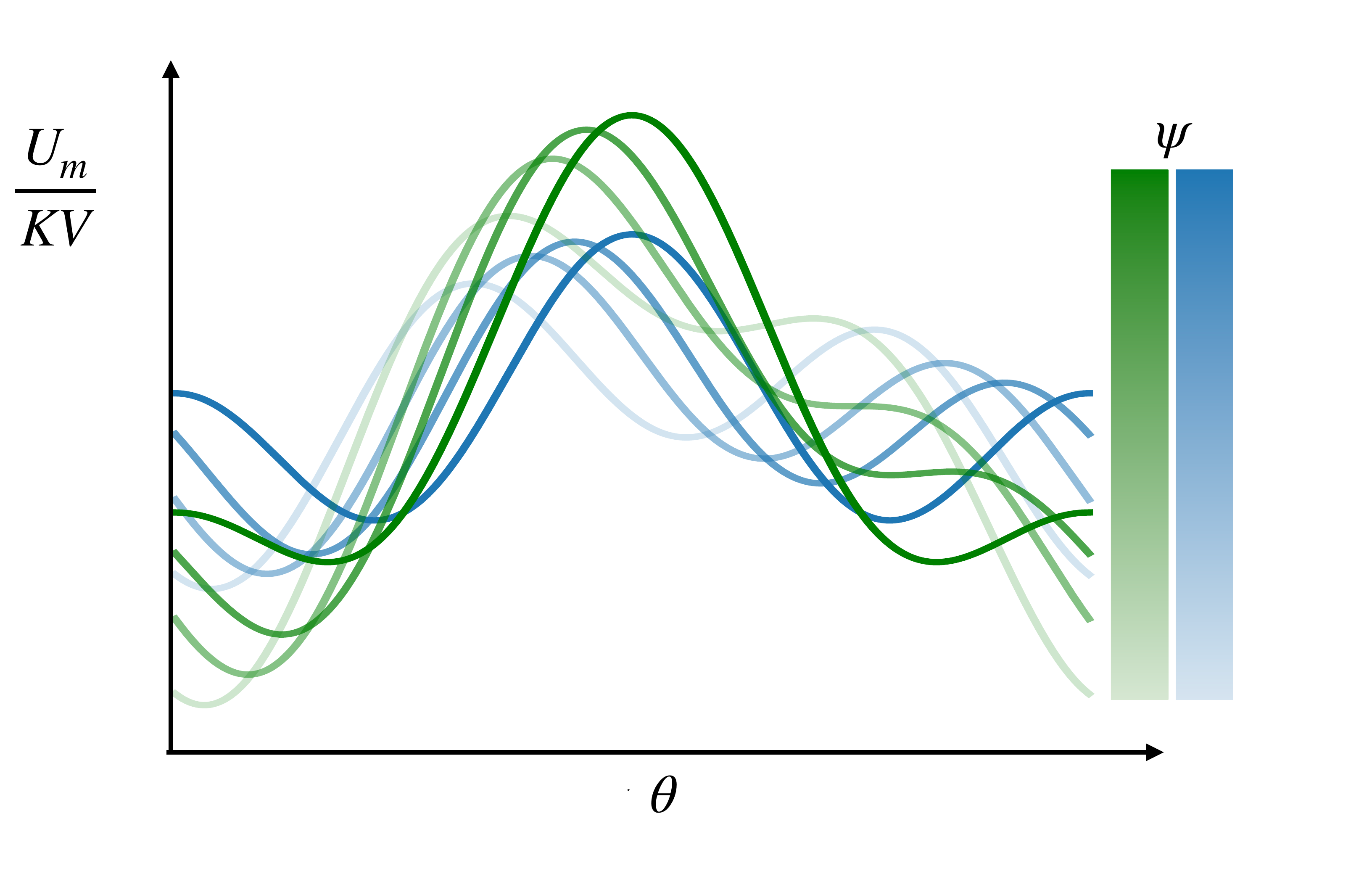}
 \caption{Visualization of the energy (density) landscape in the Stoner-Wohlfarth model, as a function of $\theta$. Lines with a different color saturation correspond to energy densities for a different $\psi$. The more saturated a color appears, the more aligned the anisotropy axis is with the applied field (smaller $\psi$). Different colors correspond to different applied field strengths, where green energy density profiles correspond to a higher applied field strength then blue energy density profiles.) }
 \label{fig:doodle_new}
 \end{figure}

\noindent
The problem has three characteristic energy scales: the Zeeman energy, $\mu_0 \mu H$, the anisotropy energy, $KV$, and thermal energy, $k_B T$. The interplay between the characteristic energy scales is described in terms of two dimensionless parameters, namely the Langevin parameter:
\begin{equation}
\xi = \frac{\mu_0 \mu H}{k_B T},
\end{equation}
and the anisotropy parameter:
\begin{equation}
\sigma = \frac{KV}{k_B T}.
\end{equation}
Additionally, the interplay between anisotropy and external field is traditionally described with the dimensionless field:
\begin{equation}
    h = \frac{\xi}{2 \sigma} = \frac{\mu_0 M_s H}{2 K} = \frac{H}{H_{ani}}
\end{equation}
where $H_{ani} = 2K/\mu_0 M_s$ is the anisotropy field. For the geometry sketched in Fig.~\ref{fig:doodle}, Eq.~\ref{eq:sw_mag_ene} can be rewritten as:
\begin{equation}\label{eq::sw_minim}
    \frac{U}{KV} = - \left(\cos (\phi -\zeta) \sin \theta \sin \psi + \cos \theta \cos \psi \right)^2 - 2 h \cos \theta.
\end{equation}
where, according to~\citet{stoner1948mechanism}, global minima of the potential always lie in the plane spanned by $\vec{H}$ and $\vec{n}$ ($\phi = \zeta$). Eq.~\ref{eq::sw_minim} has two minima below the critical field strength:
\begin{equation}\label{eq:hcr}
    h_{cr}(\psi) = (\sin^{2/3}\psi + \cos^{2/3}\psi)^{-3/2}. 
\end{equation}
Above the critical field, only one global minimum remains. The particle dipole moment follows the position of the local energy minimum and instantaneously changes its direction depending on the strength of $h$ in relation to $h_{cr}$. Transitions between the energy minima are not possible below $h_{cr}$ (at $T=0$). A visualization of the energy landscape defined with Eq. \ref{eq::sw_minim} ($\phi = \zeta$), along with its dependence on $\theta$,$\psi$ and $h$, is provided in Fig.\ref{fig:doodle_new}. The simplest \emph{thermal} formulation of the total magnetic energy would include the Néel relaxation mechanism, understood as the internal relaxation achieved through thermal fluctuation induced transitions of the dipole moment between the local energy minima in the total magnetic energy of a particle. This intuitive notion of the Néel mechanism involves several important approximations, discussed at length below. Brownian relaxation instead refers to the rotational diffusion of the particle body. 

\subsection{Relevant time scales}\label{ssec::time}

Here, we discuss the timescales of the relaxation mechanisms characteristic for ensembles of magnetic nanoparticles, starting with the ones associated with internal degrees of freedom. In the absence of thermal fluctuations and an applied field, the main timescale that characterizes the magnetization relaxation in immobilized ensembles is the damping time of the Larmor precession. From the standard Landau-Lifshitz-Gilbert (LLG) equation, it can be estimated as~\cite{poperechny2014dynamic}:
\begin{equation}
    \tau_0 = \frac{1}{\alpha \omega_L} = \frac{(1+\alpha^2)M_s}{2 \alpha \gamma K},
\end{equation}
where $\omega_L = \gamma \mu_0 H_{ani}/(1 + \alpha^2)$ is the precession frequency (in the absence of an applied field), $\alpha$ is the phenomenological damping parameter with typical values in the range between 0.01 and 0.1,
$\gamma \simeq 1.76 \times 10^{11}~\textrm{s}^{-1}\textrm{T}^{-1}$ is the gyromagnetic ratio.

At any non-zero temperature, the magnetic moment is subjected to thermal fluctuations. 
The characteristic time scale of its rotational diffusion in the limit of weak field and weak anisotropy (\textit{i.e.}, at  $\xi \ll 1$  and  $\sigma \ll 1$) is given by
\begin{equation}
\tau_D = \frac{(1 + \alpha^2)\mu}{2 \alpha \gamma k_B T} = \sigma \tau_0.
\end{equation}
The time scale $\tau_D$ (sometimes referred to as the Debye time) arises naturally, as thermal noise is introduced into the LLG equation and it is required to reproduce the Gibbs-Boltzmann distribution in thermodynamic equilibrium~\cite{garcia1998langevin}.

However, for nanoparticles with $\sigma \ge 1$, it is common to distinguish two mechanisms -- 1) jumps between energy minima and 2) fast fluctuations in the vicinity of a given minimum~\cite{raikher2004nonlinear,coffey2012thermal}. In~\citet{raikher2004nonlinear} terminology, these are called ``interwell'' and ``intrawell'' relaxation processes, respectively. For sufficiently large anisotropies, $\sigma \gg 1$, one might neglect the intrawell processes entirely. 
In this approximation, magnetic moments are always in one of the local minima and are only allowed to jump between them. 
L. N\'eel in 1949 was the first to use this approximation and suggest that the magnetization relaxation time due to over-barrier thermal jumps follows an Arrhenius-like law~\cite{neel1949influence}.
In 1963, the N\'eel result was improved by W.F. Brown Jr.~\cite{brown1963}.
With the help of the Kramers escape rate theory, Brown gave the following high-barrier approximation for the magnetization relaxation time (the so-called N\'eel time):
\begin{equation}~\label{eq::tn_brown}
\tau_N(\sigma \gg 1) = \frac{\tau_D}{2\sigma}\sqrt{\frac{\pi}{\sigma}}\exp \sigma. 
\end{equation} 

Formally, the Néel time $\tau_N$ is the inverse of the smallest non-vanishing eigenvalue of the Fokker-Planck equation, that Brown formulated on the basis of the stochastic LLG equation. 
Eq. \ref{eq::tn_brown} is only valid for uniaxial particles in zero field. 
However, in the literature one can find multiple approximations for $\tau_N$ in static fields of different orientations, as well as for other anisotropy types~\cite{coffey2012thermal,chalifour2021magnetic}. 
Brown, in his work, also demonstrated that the zero-field rate of over-barrier jumps is not affected by the Larmor precession and that the corresponding term in the Fokker-Planck equation can be safely omitted when one is looking for $\tau_N$.
The same is true if the field is applied parallel to the easy axis ($\psi = 0$). This is a general property of the Fokker-Plank-Brown equation for problems that possess axial symmetry. 
However, for all other cases ($\psi \neq 0$ or different anisotropy types), precession will affect magnetic relaxation and should in principle be explicitly taken into consideration~\cite{raikher1975theory,kalmykov1997}. 
Omitting precession in the general case limits the results to the field frequencies below the ferromagnetic resonance range.

Finally, we introduce the timescales of relaxation mechanisms associated with external degrees of freedom in nanoparticles. If the particle is suspended in a viscous medium, two more relaxation times, connected to its mechanical rotation, must be introduced~\cite{rauikher1994effective}. If a particle is coated with a non-magnetic shell of width $l$, its hydrodynamic diameter can be calculated as $d_H = d + 2l$. The inertial decay time is:
\begin{equation}
    \tau_J = \frac{J}{\Gamma_R} = \frac{\rho d^5}{60 \eta d_H^3}, 
\end{equation}
where $J = 0.1 \rho V d^2$ is the moment of inertia, $\rho$ is the particle material density (here we assume that all the mass is concentrated in the magnetic core), $\Gamma_R = \pi \eta d^3_H$ is the rotational friction coefficient, $\eta$ is the fluid viscosity.
\noindent
The rotational Brownian time is:
\begin{equation}
    \tau_B = \frac{\Gamma_R}{2 k_B T} = \frac{\pi \eta d_H^3}{2 k_B T}.
\end{equation}


\subsection{Implementation details}\label{ssec:model}

Here we outline our tSW algorithm, specifically, how we simulate the internal dynamics and the Néel relaxation mechanism in magnetic colloids. For more details on the simulation approach, we refer the reader to Section~\ref{sec:method}. The algorithm requires $\mu$, $h$, and $\sigma$ as input parameters. The algorithm can be logically separated into three steps:
\begin{enumerate}
    \item Finding the extrema in Eq.~\ref{eq::sw_minim}, for the current state of the particle.
    \item Calculation of the energy barrier to estimate the transition probability.
    \item Updating the dipole moment orientation based on a trial move against the transition probability. 
\end{enumerate}
Firstly, $h_{cr}$ (Eq.~\ref{eq:hcr}) is calculated taking the current angle between the vector $h$ and the particle anisotropy axis at its position, $\phi$. For a given $\phi$, the algorithm finds $\theta$ which minimizes the total magnetic energy (Eq.~\ref{eq::sw_minim}) $\theta^{\prime}_{min}$, which is closest to the previous dipole moment state. This is ensured by initializing the state of the energy minimizer with the previous particle state. If the field acting on the particle is less than $h_{cr}$, the algorithm proceeds to find a $\theta$ that maximizes the total magnetic energy from both sides of $\theta^{\prime}_{min}$, denoted with $\theta^{\prime}_{max}$ and $\theta^{\prime \prime}_{max}$. With this information, the algorithm calculates the energy barriers found on both sides of $\theta^{\prime}_{min}$, 
\begin{equation}
\begin{aligned}
    \Delta E^{\prime}&=\dfrac{1}{KV}|U(\phi,\theta^{\prime}_{max})-U(\phi,\theta^{\prime}_{min})|, \\
\Delta E^{\prime \prime}&=\dfrac{1}{KV}|U(\phi,\theta^{\prime \prime}_{max})-U(\phi,\theta^{\prime}_{min})|.
\end{aligned}
\end{equation}
and selects the smaller energy barrier $\Delta E=\min(\Delta E^{\prime}, \Delta E^{\prime \prime})$ to estimate a characteristic timescale of the Néel relaxation process, using a modified form of Eq.~\ref{eq::tn_brown}:
\begin{equation}
    \tau_N = \frac{\tau_D}{2\sigma}\sqrt{\frac{\pi}{\sigma}} e^{\Delta E\sigma},
\end{equation}
and from it, the transition probability $p=1-e^{\delta t/\tau_N}$, where $\delta t$ is the integration step time~\cite{chantrell2000calculations}. It is suggested in the literature, for example, in \citet{chuev2007nanomagnetism}, that one should estimate the transition probability as an average, considering both $\Delta E^{\prime}$ and $\Delta E^{\prime \prime}$. While the general mean barrier approach is a topic for discussion~\cite{koraltan2020dependence}, in our testing, we did not find a measurable difference between the two approaches and used the simpler one that is compatible with the analytical formulation of tSW. We make a trial move by casting a random number and comparing it against the transition probability, akin to a Metropolis step. If the trial move is successful, the algorithm finds a new $\theta$ that minimizes Eq.~\ref{eq::sw_minim}, denoted by $\theta^{\prime \prime}_{min}$, and sets the dipole moment to point in accordance with the new minimum. Alternatively, the dipole moment is set to point in accordance with $\theta^{\prime}_{min}$.
\begin{figure}[t]
    \centering
    \includegraphics[width=\linewidth]{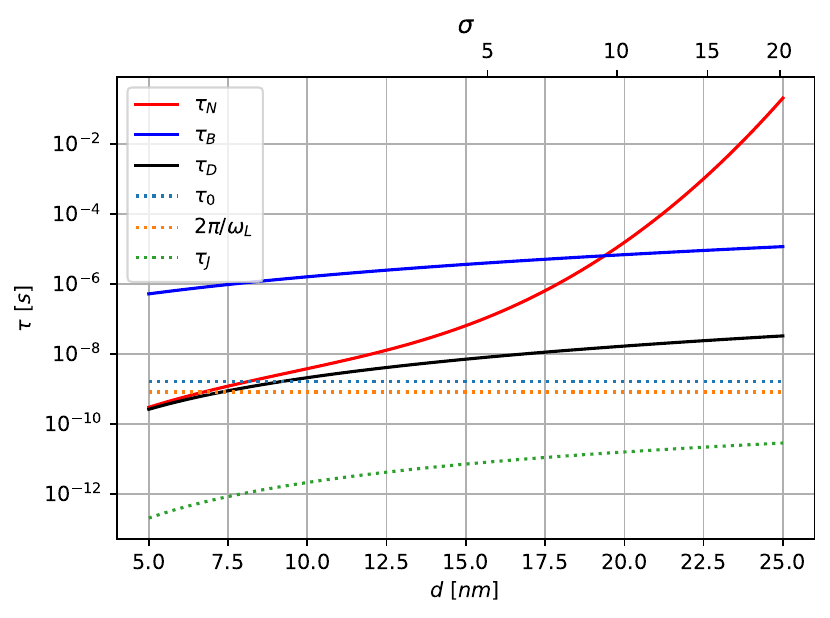}
    \caption{Characteristic timescales involved in relaxation processes associated with ensembles of magnetic nanoparticles. Timescales are calculated for magnetite ($M_s=480\; kA/m; K=10\; kJ/m^3$), as a function of particle size and temperature, where we choose $\alpha=0.08$. The top x-axis is showing what $\sigma$ to qualify particle size (bottom x-axis) with respect to temperature (not shown).}
    \label{fig:timescales}
\end{figure}
\section{Results and Discussion}\label{sec:results}
Looking at Fig.~\ref{fig:timescales}, one can get a sense of the considerations and constraints encountered in simulation studies of magnetic soft matter, where the objective is often to scrutinize collective properties at long timescales. We plot the dependence of Ne\'el ($\tau_N$), Brownian ($\tau_B$), internal diffusion ($\tau_D$), period of the precession ($2\pi/\omega_L$), precession damping ($\tau_0$) and inertial decay ($\tau_j$) times as functions of magnetite particle diameter $d$. The range of the anisotropy constant $\sigma$ between $5-20$ covers the most commonly studied nanoparticle sizes and materials. If one was to simulate the magnetodynamics of the nanoparticles explicitly, accounting for all their characteristic magnetic relaxation timescales by, for example, using the aforementioned LLG equation~\cite{landau1992theory,gilbert2004phenomenological,helbig2023self}, it becomes apparent that $\tau_J$ is not small enough to be safely ignored. This means that one must resolve timescales less than \emph{picoseconds} to be able to include full magnetodynamics in simulation. This is prohibitively low for any simulation study concerned with bulk properties. So, from practical considerations, one would like to simulate timescales where high frequency processes ($\leq 10^{-8}$) can be safely ignored. However, a competition remains between $\tau_N$ and $\tau_D$ on the internal relaxation, and $\tau_B$ on the external  relaxation side. Specifically, within the $5<\sigma<20$ range, there are regions where $\tau_N$ and $\tau_D$ are comparable within simulation time, to $\tau_N$ becoming prohibitively large, if one was to consider relaxation associated with $\tau_D$. Understanding the optimal approach to be able to include relevant magnetodynamic processes in this range, is precisely the question we address below.

\begin{figure*}[t]
    \centering
    \includegraphics[width=\linewidth]{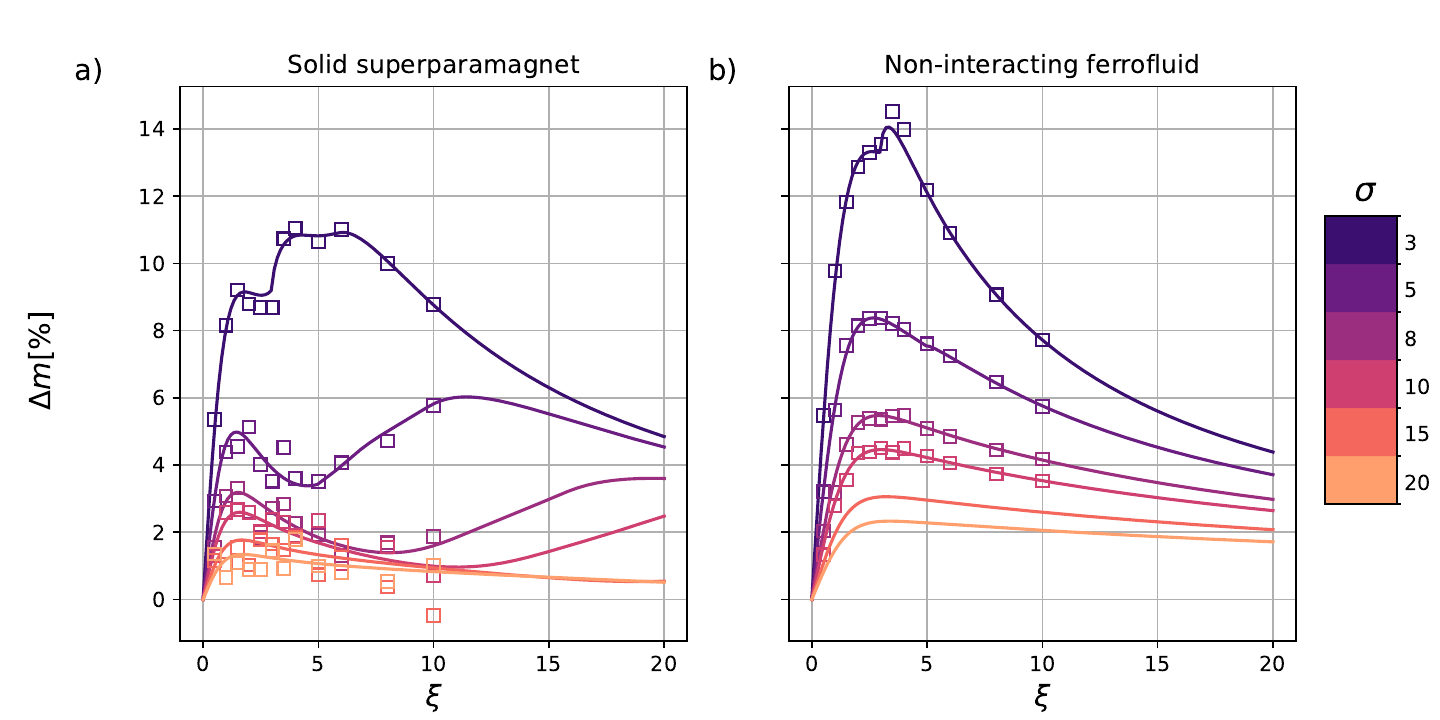}
   \caption{The absolute magnetization error $\Delta m$ (in percent), as a function of $\xi$, for a non-interacting (a) solid superparamagnet, where the analytical tSW results (solid lines, given by Eq.~\ref{eq::m_hws}) and simulations (symbols) are compared to the continuum-level description (given by Eq.~\ref{eq:master_solid}); (b) ferrofluid, where the analytical tSW results (solid lines, given by Eq.~\ref{eq:mtSW_ferrofluid}) and simulations (symbols) are compared to the continuum-level description (given by Eq.~\ref{eq:langevin}). The color bar indicates the anisotropy parameter $\sigma$ values explored.}
    \label{fig:master_mag}
\end{figure*}

\subsection{Static Properties}\label{ssec::equil}

At a finite temperature, an ensemble of magnetic colloids in a static homogeneous magnetic field reaches thermodynamic equilibrium at times larger than the system’s characteristic relaxation time. For any particular configuration of anisotropy axes in a nanoparticle ensemble, the equilibrium state is uniquely defined by ($\sigma$, $\xi$). In Fig.~\ref{fig:master_mag}, we scrutinize the tSW model in its ability to correctly reproduce equilibrium magnetization, $m$, of an ensemble of magnetic nanoparticles. The equilibrium magnetization of the system is the total sum of the magnetic moments $\mu$, projected onto the axis of the applied magnetic field, and normalized to unity throughout the manuscript. As previously mentioned, we distinguish cases where the particles exhibit either only Néel (solid superparamagnet) or both Brownian and Néel (ferrofluid) relaxation mechanisms. 

Within the tSW model, the discrete states in which the magnetic moment can be for a \textit{solid superparamagnet}, correspond to the energy minima in the total magnetic energy (Eq.~\ref{eq:sw_mag_ene}). The equilibrium magnetization can be obtained if one performs a summation over a set of allowed discrete orientations:
\begin{equation}
    m_H^{tSW} = \frac{\sum_i \cos \theta_{min, i} \exp \big[ - U(\theta_{min, i}) / k_B T \big]}{\sum_i \exp \big[ - U(\theta_{min, i}) / k_B T \big] },
\end{equation}
where $\theta_{min,i}$ are orientations minimizing the energy at given values of $\xi$, $\sigma$ and $\psi$, $i = 1,2$ is the number of a corresponding minimum. 
To get the magnetization of an ensemble with randomly distributed anisotropy axes, one needs to average $m_H^{tSW}$ over all possible orientations of $\vec{n}$:
\begin{equation}\label{eq::m_hws}
    m^{tSW} = \int^{\pi/2}_0 m_H^{tSW}(\xi, \sigma, \psi) \sin \psi  \text{d} \psi.
\end{equation}
The integral in Eq.~\ref{eq::m_hws} can be calculated if we consider components parallel and perpendicular to the applied field direction separately ($h < 1$), $m^{tSW}_{\parallel}$ and $m^{tSW}_{\perp}$, respectively, where:
\begin{equation}
\begin{aligned}
    m^{tSW}_{\parallel} &= \tanh \xi, \\
m^{tSW}_{\perp} &= \xi / 2 \sigma.
\end{aligned}
\end{equation}
The problem of describing the equilibrium magnetization of an ensemble of immobilized, non-interacting, randomly oriented uniaxial particles, has been solved based on a generalization of the Langevin function for the case of solid dispersions with random orientation texture~\cite{kuznetsov2018equilibrium,williams1993superparamagnetism,cregg1999series}. The full expression together with an outline of the derivation are provided in the Appendix A. 

In the case of a \textit{non-interacting ferrofluid}, $\vec{n}$ and $\vec{e}$ are no longer independent variables - in the tSW model, there exists a unique set of allowed magnetic moment orientations, for any given orientation of the anisotropy axis. The equilibrium magnetization can be obtained in a similar form to Eq.~\ref{eq::m_hws}, where:
\begin{widetext}
\begin{equation}\label{eq:mtSW_ferrofluid}
    \begin{aligned}
         m^{tSW} = \int^{\pi/2}_0 m_H^{tSW}(\xi, \sigma, \psi) f_{n}^{tSW}(\xi, \sigma,\psi)\sin \psi  \text{d} \psi,\\ f_{n}^{tSW}(\xi, \sigma,\psi) = \frac{\sum_i \exp\big[ - U(\theta_{min, i}(\psi)) / k_B T \big] }{\int^{\pi/2}_0 \Big(\sum_i \exp \big[ - U(\theta_{min, i}(\psi)) / k_B T \big] \Big) \sin\psi  \text{d} \psi}.
    \end{aligned}
\end{equation}
\end{widetext}
For $\xi = 0$, the distribution is random and $f_{n}^{tSW} = 1$. In general, however, the magnetization of a non-interacting ferrofluid is given by the Langevin function~\cite{elfimova2019static}:
\begin{equation}\label{eq:langevin}
m^{L} = mL(\xi), \ \ L(\xi) = \coth(\xi) - \frac{1}{\xi}. 
\end{equation}
In this case, the magnetization does not depend on the anisotropy parameter $\sigma$.

In Fig.~\ref{fig:master_mag}, we are showing the absolute magnetization error, $\Delta m=100\cdot |m^{tSW}-m^{cont}|/m$, as a function of $\xi$. Note that $m^{tSW}$ represents either the simulated data (symbols) or the analytical results based on the tSW model (lines). Specifically, $m^{tSW}$ corresponds to  Eq.~\ref{eq::m_hws} for a solid superparamagnet and Eq.~\ref{eq:mtSW_ferrofluid} for a ferrofluid. Similarly, $m^{cont}$, representing the continuum-level description, is given by Eq.~\ref{eq:master_solid} for a solid superparamagnet and Eq.~\ref{eq:langevin} for a ferrofluid. Fig.~\ref{fig:master_mag} demonstrates that the tSW model can reproduce the equilibrium magnetization within a few percent of the ground truth, in all cases except for particle sizes and/or temperatures where $\sigma \leq 5$, where one can expect a purely super-paramagnetic response. Below this $\sigma$ range, we encounter the limitations of the model. Let us address these limitations. In a solid superparamagnet, the only relaxation mechanism that the particles can leverage to reach equilibrium is Néel.

In the tSW model, there is only a single state in which the dipole moment can be above $h_{cr}$. In other words, the tSW model is equivalent to the classical SW model above $h_{cr}$. In general, however, the magnetic moment orientation can fluctuate relative to the states allowed in the classical SW model~\cite{kuznetsov2018equilibrium,williams1993superparamagnetism,cregg1999series}. The frequency of these intrawell fluctuations is inversely proportional to $\sigma$. With this in mind, one can understand why the tSW model overestimates the equilibrium magnetization of low $\sigma$ particles in the large $\xi$ region. Moreover, the intrawell modes matter for high $\xi$ regardless of $\sigma$, as Néel relaxation ($1/p_1$) is no longer sufficient for an accurate description of magnetic relaxation~\cite{garanin1997fokker,poperechny2023multipeak}. Hence, we observe regions with a slight increase in $\Delta m$ with growing $\xi$. Finally, the importance of intrawell states relative to interwell states increases with decreasing $\xi$, since the applied field is not strong enough to constrain the dipole moment fluctuations perpendicular to the anisotropy axis.

Moving on to the non-interacting ferrofluid system, where both Néel and Brownian relaxation are at play, the same conclusions largely hold as for a solid superparamagnet. However, one can note a marginally worse agreement between the tSW model and the continuum-level description. Consider the high $\xi$ region, where one can assume Néel relaxation plays a minimal role. For high $\sigma$, we have seen that for a fixed random distribution of the anisotropy axis, the SW reproduces the equilibrium state very well. In the same circumstances, given a non-random distribution of anisotropy axes, the relative error is almost quadrupled. The difference in relative error between the left and right subplots in Fig.~\ref{fig:master_mag} is related to the fact that, for a solid superparamagnet, the anisotropy axis orientations are randomly distributed and fixed, whereas for a ferrofluid, this is not the case. The classical SW model estimates the equilibrium magnetization incorrectly, but not equally incorrectly for all anisotropy axis distributions. In fact, it can be inferred by comparing the left and right side subplots in Fig.~\ref{fig:master_mag}, that the more the anisotropy axes are aligned with the magnetic field, the worse is the equilibrium state estimation in the classical SW model.  

Finally, it is worth highlighting a particularity of the tSW model, related to the initial susceptibility for low $\sigma$. Initial magnetic susceptibility is an important characteristic of the material’s linear response to a weak applied field (or ``probing field"):
\begin{equation}
    \chi = \frac{\partial m}{\partial H} \Big|_{H \rightarrow 0}.
\end{equation}
It can be shown that in the tSW model, the total static susceptibility is given by:
\begin{equation}\label{eq:static_susc}
\chi^{tSW}(0) = \chi_L (1 + 1/\sigma),
\end{equation}
\noindent
where $\chi_L$ is the static susceptibility of an isotropic superparamagnet (Langevin susceptibility):
\begin{equation}
\chi_L = \frac{\mu_0 \mu^2 N}{3 V k_B T}.
\end{equation}
In the limit of no particle anisotropy, the static susceptibility should correspond to the Langevin susceptibility~\cite{elfimova2019static}. However, this is not the case in the tSW model, where the values diverge. Although the tSW model reproduces the characteristic internal relaxation time related with Néel relaxation well, it is inadequate for simulations of small enough nanoparticles, or at high enough temperatures where $\sigma<5$.

Having addressed the limitation of the tSW at length, it is important to note that the absolute error for $\sigma \geq 5$ remains less than $10\%$, which qualifies the tSW approach as surprisingly accurate within a wide ($\sigma,\xi$) range of applicability. It is apparent that the dominant internal relaxation mechanism is the Néel one, even for relatively small $\sigma$. Furthermore, our implementation of the tSW is in agreement with the analytical prediction for the model across the board. The key insight here is that the validity of the tSW model is most constrained by the limitations of the classical SW model, which are well understood and documented. The discrepancies stemming from the lack of intrawell modes are less important, and we do not see this as a substantial additional limitation to the model. Moreover, as long as one is considering non-interacting systems, one can compensate for the overestimation stemming from the classical SW model by tuning the saturation magnetization of the colloids to fit the correct equilibrium magnetization curves. 

\begin{figure}
    \centering
    \includegraphics[width=\linewidth]{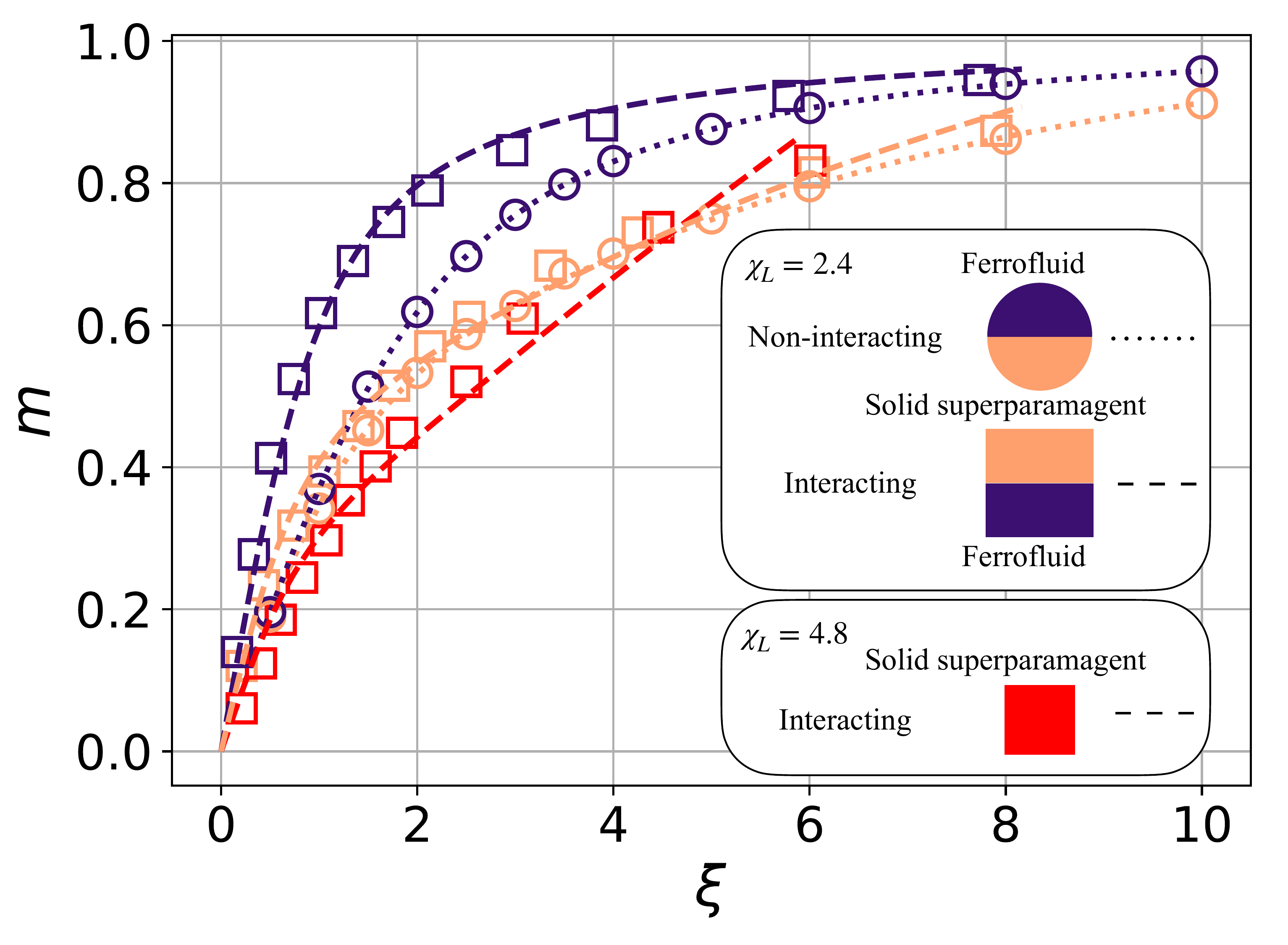}
    \caption{
    The magnetization $m$ as a function of $\xi$, for a solid superparamagnet (orange for $\chi_L=2.4$, red for for $\chi_L=4.8$) and a ferrofluid (violet). Interacting systems are represented by square symbols (simulation data) and dashed lines (theory), while non-interacting systems use circles (simulation data) and dotted lines (theory). Dotted lines correspond to Eqs.~\ref{eq:mtSW_ferrofluid} and~\ref{eq::m_hws} (ferrofluid and solid superparamagnet, respectively). Dashed lines represent MMTF extensions of these equations, with an additional scaling factor $\alpha(\xi)=a+b\xi$ for the solid superparamagnet (details in text) where $(a,b)$ are fitted parameters (for $\chi_L=2.4$, $a=0.8151$ and $b = 0.0213$; $\chi_L=4.8$, $a=0.4989$ and $b = 0.0778$). Interacting systems were simulated at $\lambda = 3$. The effects of demagnetization fields in have been removed from the simulation data.}
    \label{fig:interactions}
\end{figure}

\begin{figure*}
    \centering
    \includegraphics[width=0.8\linewidth]{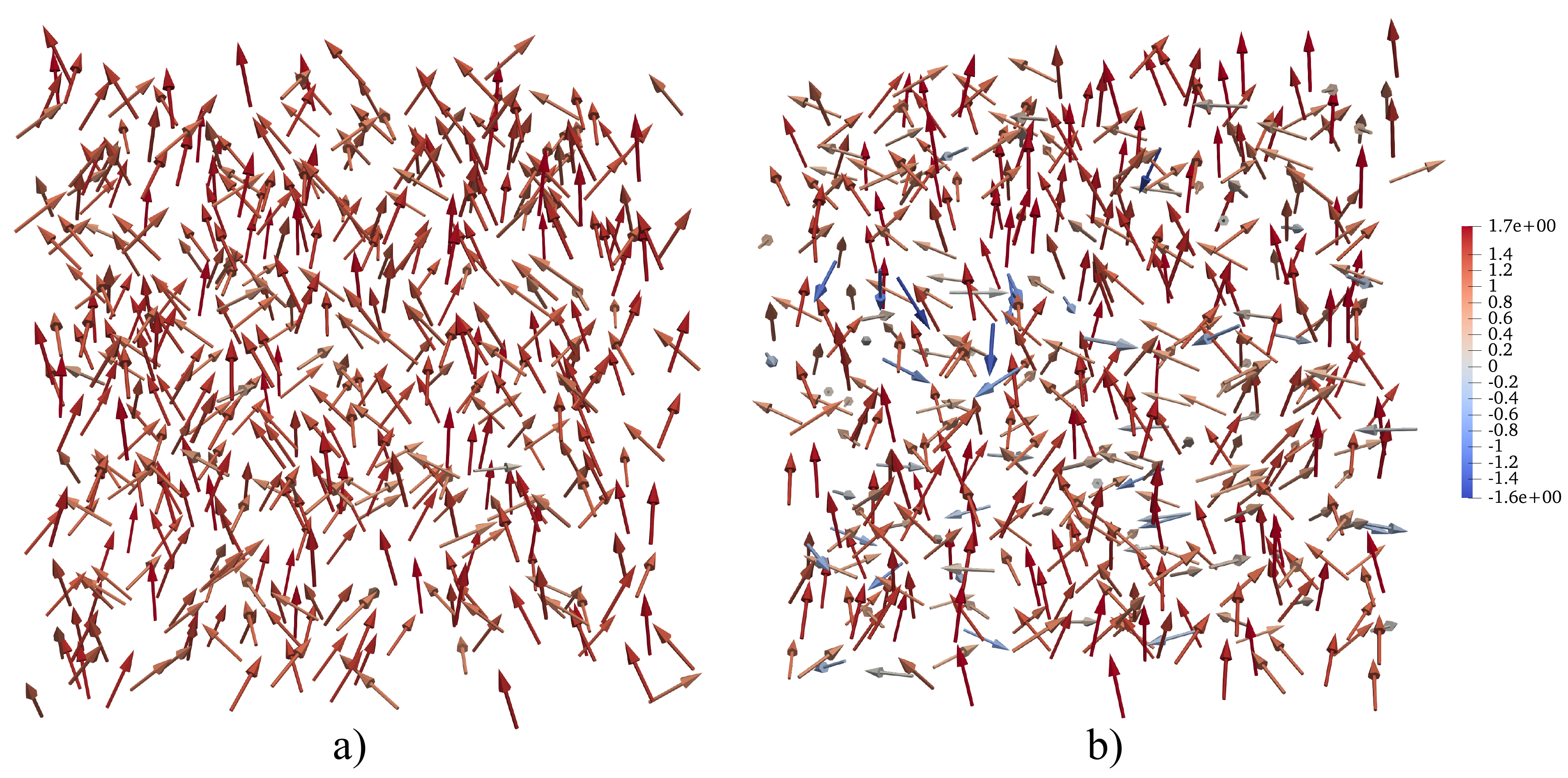}
    \caption{
    Visualization of the dipole moment orientations in: (a) non-interacting; (b) interacting solid superparamagnet ($\chi_L=4.8$), at $\xi=5$. The magnetic field is applied along the $z$-axis. Color-coding corresponds to the $z$-component of the dipole moment, as indicated by the color bar. The arrow size of the dipole moments is exaggerated for visualization purposes and has no physical meaning.
    }
    \label{fig:interactions_paraview}
\end{figure*}

The tSW model can also be used to study interacting systems. From the point of view of the model, there is no difference between interacting and non-interacting systems. The model requires a magnetic field as an input parameter. If one is studying interacting systems, it is necessary to pass an effective field (a vectorial sum of the total dipole field the nanoparticle feels and the applied magnetic field) to the solver, instead of the applied magnetic field. All other aspects of dipole-dipole interactions are exogenous to the model and must be handled separately. In Fig.\ref{fig:interactions}, we consider how dipole-dipole interactions affect the equilibrium magnetization for $\sigma=5$ solid superparamagnet and ferrofluid systems, respectively. For the parameters we used, a $\sigma=5$ nanoparticle corresponds to the dipolar coupling parameter $\lambda=3$.
The $\lambda$ parameter quantifies the strength of the dipole-dipole interaction with respect to thermal fluctuations, when two dipoles are collinear and at a touching distance, and is given by:
\begin{equation}
    \lambda=\frac{\mu_0 \mu^2}{4\pi d^3k_bT}.
\end{equation}
While the magnetization of non-interacting ferrofluids is known to follow the Eq. \ref{eq:langevin}, the magnetization of interacting ferrofluids is well described with the modified mean-field theory (MMFT) instead, originally presented in \citet{ivanov2001magnetic}, as long as dipolar forces do not lead to significant cluster formation \cite{2007-ivanov}. The first-order MMFT expansion of the Langevin parameter $\xi_e =\xi + \chi_L m^{L}(\xi)$, leads to the following magnetization: $m=m^{L}(\xi_e)$. We applied an MMFT correction to Eq.\ref{eq:mtSW_ferrofluid}, to obtain an analytical expression of an interacting ferrofluid within the scope of the tSW model (purple dashed line). This simple treatment is in good agreement with our simulated data for an interacting ferrofluid (purple square symbols). We reproduce the expected behavior, which is that dipole-dipole interactions tend to increase the magnetic response in a ferrofluid. However, it must be said that it is a non-trivial claim that the MMFT approach should work for a two-state model such as the tSW model, and a more rigorous derivation would be necessary. Comparing the simulated data for non-interacting and interacting solid superparamagnet systems at $\chi_L=2.4$ (orange circles and square symbols, respectively), we see that interactions lead to only a slight increase in the magnetic response, that differs qualitatively from what one might expect from \citet{elfimova2019static}. For $\chi_L=4.8$ (red square symbols), we can note pronounced decrease in magnetization in the low to moderately high $\xi$ range. One can get an intuitive sense of what could be the mechanism behind these results by looking at Fig. \ref{fig:interactions_paraview}. Comparing \ref{fig:interactions_paraview}(a) and \ref{fig:interactions_paraview}(b), it should be visually accessible that the dipoles in a non-interacting system are more aligned with the applied field than for an interacting system. In an interacting system, the dipoles are coupling to a local field, which is a complex function of the dipolar fields and the applied field. In the case of an interacting solid superparamagnet, our conjecture is that dipole-dipole interaction correlated vortexes are formed, which impedes the magnetic response of the system. The mechanism is reminiscent of the one of ring formation in strongly interacting magnetic fluids \cite{2013-kantorovich-prl} and vortex formation in magnetic multicore particles \cite{kuznetsov2023multicore,solovyova2023orientation}. In analogy to these works, we consider these vortices as "magnetic holes" that can be effectively subtracted from the system’s magnetic response. The functional dependence of the fraction of magnetic holes on the applied magnetic field strength is unknown and a matter for a separate, in-depth study. For the purposes of this work, we assumed that the fraction of magnetic holes decreases linearly with an applied magnetic field strength. With this assumption, we can incorporate dipole interactions using an MMFT based extension to Eq. \ref{eq::m_hws} to obtain an analytical form for the magnetization for an interacting solid superparamagnet, shown in Fig.\ref{fig:interactions}, in the following form: $m^{tSW} = \alpha(\xi)m^{tSW}(\xi_e)$, where $\xi_e = \xi + \chi_L \cdot \alpha(\xi) \cdot m^{tSW}(\xi)$, and $\alpha$ is a linear function of $\xi$. We see a rather good agreement between the analytical expression (orange and red dashed line, for $\chi_L=2.4$ and $\chi_L=4.8$, respectively) and the simulated data (orange and red square symbols, for $\chi_L=2.4$ and $\chi_L=4.8$, respectively). The results shown in Fig. \ref{fig:interactions} demonstrate the utility of the tSW approach in the study of interacting systems, and hint at the promise of similar highly scalable methods to open new horizons in the study of magnetic soft matter.

In summary, the tSW model is a reasonably accurate approach to introduce internal magnetization dynamics for $\sigma \geq 5$ nanoparticles via Néel relaxation, that can reproduce the correct equilibrium in this $\sigma$ range. Having established that, the discussion can proceed to the study of dynamics, which is the type of study where the tSW model can benefit researchers interested in long timescales and bulk-sized systems.

\subsection{Dynamic Properties}\label{ssec::susc}

One of the key collective properties of any magnetic soft matter system is its characteristic magnetic relaxation time. This is especially relevant for systems designed for heating applications. The magnetic relaxation time can be assessed by calculating the dynamic susceptibility spectra and locating the maximum/maxima of the susceptibility’s imaginary part.
\begin{figure*}[t]
    \centering
    \includegraphics[width=\linewidth]{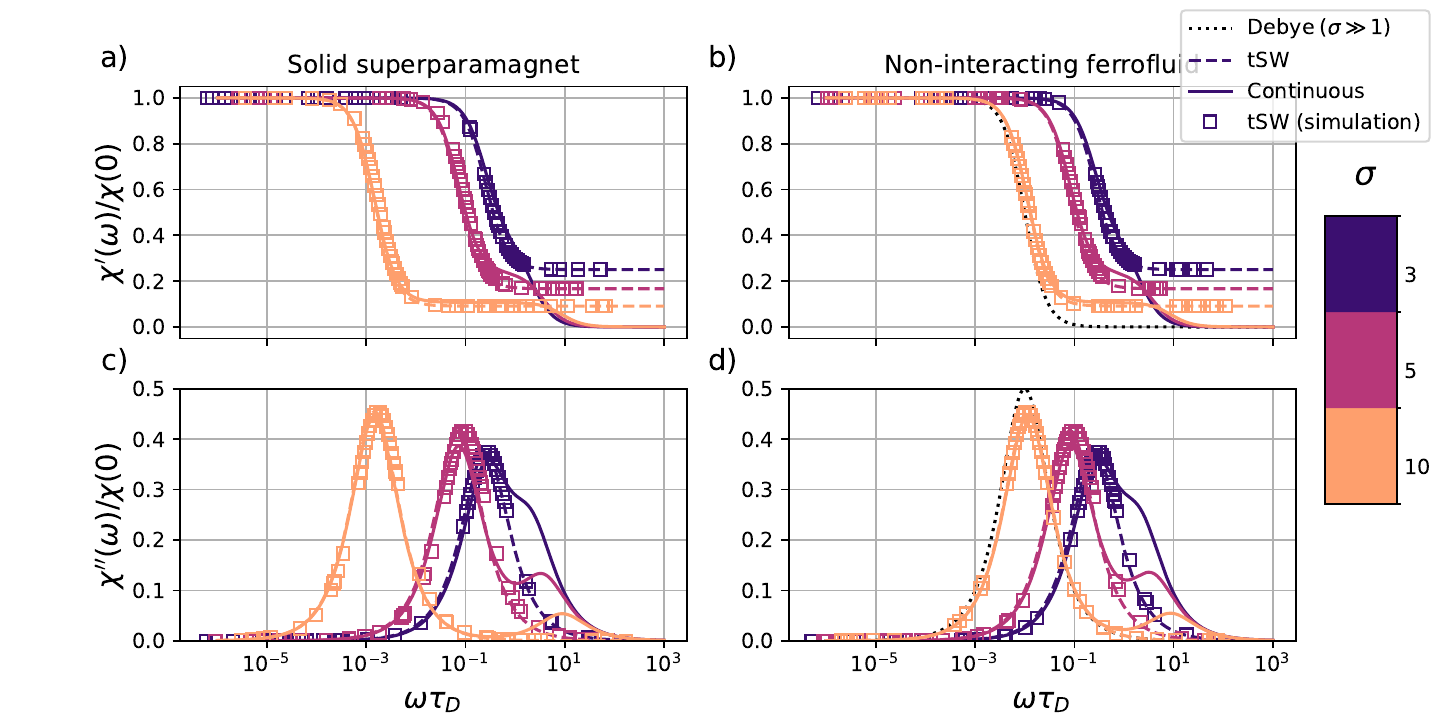}
    \caption{Dynamic susceptibilities as a function of $\omega \tau_D$, where $\tau_D/\tau_B=0.02$, for a non-interacting (i) solid superparamagnet ((a) real part; (c) imaginary part), where the analytical tSW results (dashed lines) and simulations (symbols) are compared to the continuum-level description (solid lines); ferrofluid ((b) real part; (d) imaginary part), where the analytical tSW results (dashed lines) and simulations (symbols) are compared to the continuum-level description (solid lines). The analytical tSW corresponds to Eq.~\ref{eq::susc_tSW_master}, where the input relaxation time is either the Néel time, or the effective time $\left(\tau^{-1}_{N} + \tau_B^{-1}\right)^{-1}$ (solid superparamagnet and ferrofluid, respectively) . The continuum-level description corresponds to Eq.~\ref{eq::sus_cont_master} where the relaxation times $\tau_{\parallel}$ and $\tau_{\perp}$ are given either by Eq.~\ref{eq::sus_cont_solid} or Eq.~\ref{eq::sus_cont_ferro} (solid superparamagnet and ferrofluid, respectively). The color bar indicates the anisotropy parameter $\sigma$ values explored.}
    \label{fig:susc}
\end{figure*}
\noindent
Let us consider an ensemble of immobilized, non-interacting, randomly oriented particles with uniaxial magnetic anisotropy. In a harmonically oscillating field $H = H_0 \cos \omega t$, where $t$ is the time and $\omega$ is the oscillation frequency, the magnetization can generally be represented as a series:
\begin{equation}
    m(t) = \sum_{k = 1}^{\infty} \left(m'_k (\omega) \cos k \omega t + m''_k (\omega) \sin k \omega t \right),
\end{equation}
where $m'$ and $m''$ are the Fourier coefficients. In the linear response regime, $H_0 \rightarrow 0$, only the first harmonic, $k = 1$, remains. Here, the dynamic complex susceptibility can be introduced as
\begin{equation}\label{eq::chi_replaced}
     \chi(\omega) = \chi'(\omega) + i \chi''(\omega), 
\end{equation}
where:
\begin{equation}
    \chi'(\omega) = m'_1(\omega) / H_0, \:\: \chi''(\omega) = m''_1(\omega) / H_0.
\end{equation}
 For each particle, the susceptibility can be split into components parallel and perpendicular to its anisotropy axis, $\chi_{\parallel}$ and $\chi_{\perp}$, respectively. Assuming that a probing field $|\vec{H}| = H$ is in a $xz$-plane and forms an angle $\psi$ with an easy axis, the magnetization in the linear regime is given by:
\begin{equation}
    \vec{m} = (\chi_{\perp} H \sin \psi, 0, \chi_{\parallel} H \cos \psi),
\end{equation}

where, the total particle susceptibility obeys the superposition rule \cite{raikher2004nonlinear}:
\begin{equation}\label{eq::sus_cont_master}
    \chi=\left(\frac{1}{3}\chi_{\parallel} + \frac{2}{3}\chi_{\perp}\right),
\end{equation}
where, within the continuous approach and without gyromagnetic effects~\cite{raikher2004nonlinear}:
\begin{equation}
    \begin{aligned}
\chi_{\parallel}(\omega) &\simeq \frac{\chi_{\parallel}(0)}{1 - i \omega \tau_{\parallel}}, \:\:\: \chi_{\parallel}(0) = \chi_L (1 + 2 S),\\ 
\chi_{\perp}(\omega) &\simeq \frac{\chi_{\perp}(0)}{1 - i \omega \tau_{\perp}},\:\:\: \chi_{\perp}(0) = \chi_L (1 - S),
\end{aligned}
\end{equation}
the order parameter $S$ is given by:
\begin{equation}
    S =\frac{3}{4} \frac{1}{\sigma} \Bigg[ \frac{\sqrt{\sigma}}{D\big(\sqrt{\sigma}\big)} - 1 \Bigg] - \frac{1}{2},
\end{equation}
$D$ is the Dawson function:
 \begin{equation}
D(x) = e^{-x^2} \int_0^x e^{t^2} \,  \text{d}t.
 \end{equation}
This relation must be true for both the static and dynamic case. For immobilized particles: 
\begin{equation}\label{eq::sus_cont_solid}
\begin{aligned}
    \tau_{\parallel} &= \tau_D \frac{e^{\sigma} - 1}{2\sigma}\left[ \frac{1}{1 + 1/\sigma}\sqrt{\frac{\sigma}{\pi}} + 2^{-\sigma - 1}\right]^{-1},\\
    \tau_{\perp} &= 2 \tau_D \frac{1 - S}{2 + S},
    \end{aligned}
\end{equation}
In the case where particles are not immobilized, it is sufficient to replace relaxation times with effective values~\cite{shliomis1994theory}:
\begin{equation}\label{eq::sus_cont_ferro}
    \tau^*_{\parallel} = \left(\tau^{-1}_{\parallel} + \tau_B^{-1}\right)^{-1} , \\
\tau^*_{\perp} = \left(\tau^{-1}_{\perp} + \tau_B^{-1}\right)^{-1}.
\end{equation}
It is important to note that field-dependent relaxation should also be scrutinized to fully appreciate the implications from the coupling of the magnetic relaxation parallel and perpendicular to the easy axis. As is, this remains an open question for study. In the tSW model, the relaxation time $\tau_{\parallel}$ is an input parameter, which means that the dynamics of tSW parallel to the easy axis must be the same as in the continuous case. The orthogonal relaxation time $\tau_{\perp}$ is instantaneous. Therefore, the dynamic susceptibility of an ensemble of frozen, uniaxial magnetic nanoparticles  with a random distribution of anisotropy axis is given by:
\begin{equation}\label{eq::susc_tSW_master}
    \chi^{tSW}(\omega) = \frac{\chi_L}{1 - i \omega \tau_{\parallel}} + \frac{\chi_L}{\sigma}.
\end{equation}
\noindent
Once again, for a liquid, it is sufficient to introduce effective relaxation times: $\tau_{\parallel} \rightarrow \tau^*_{\parallel}$. Note that the orthogonal magnetic response (which is instantaneous and does not have any fluctuations associated with it) will not make any contribution to the zero-field magnetization autocorrelation function. Therefore, the calculated susceptibilities from the autocorrelation function when using the tSW method will be incorrect. The simulated susceptibilities have been calculated by applying a weak oscillating magnetic field (maximum amplitude corresponding to $\xi=0.1$) for a span of frequencies, in order to be able to determine the complex susceptibilities from the time evolution of the total magnetization vector (see Eq.\ref{eq::chi_replaced}). Similarly, the static susceptibilities should be estimated from the initial slope in the magnetization profiles. 

Looking at Fig.~\ref{fig:susc}, we can see that the tSW model reproduces the correct dynamics in both the solid superparamagnet and non-interacting ferrofluid systems, as quantified by the initial dynamic susceptibilities. Of course, in the initial susceptibility regime, the tSW model incorporated dipole moment fluctuations only along the anisotropy axis associated with $\tau_{\parallel}$, which in this case is equivalent to $\tau_N$. Since this is an input parameter for the model, the tSW can confidently reproduce the correct internal dynamics for the lowest frequency relaxation process ($\tau_{\parallel}$). It is apparent that Néel relaxation is very relevant even for $\sigma=10$ particles, where the dynamic susceptibility is clearly distinct from the Debye limit shown with dots. Recall that the maximum of the Debye limit is at $\omega \tau_d = 0.01$, which is set by the model as $\tau_D/\tau_N = 0.01$. The high frequency peaks, resolved in the continuum description, corresponding to $\tau_{\perp}$, cannot be captured in the tSW, simply due to the fact that the only intra-well states available to the dipole moment in the tSW model are the minima of the magnetic energy. In other words, there are no associated fluctuations perpendicular to the anisotropy axis in the $H\rightarrow 0$ regime. Outside of the initial susceptibility regime, this is not an issue as the dipole moment instantaneously follows the local energy minima, as obtained in the classical SW model, where there is always a relative angle the dipole moment takes with respect to the anisotropy axis. However, within the range of applicability we suggest ($\sigma>5$), these higher frequency processes are distinct enough that they could be averaged out while still resolving the correct dynamics. Having said that, it should also be clear that the in-field susceptibility contribution perpendicular to the anisotropy axis is frequency independent.

As suggested at the beginning of this section, the position(s) of the maxima of the imaginary part of the initial susceptibility indicate the optimal field frequencies for heating applications. Fig.~\ref{fig:susc} clearly shows that if one needs to optimize the frequency of an applied magnetic field for hyperthermia depending on the particle size and material, using a point dipole approximation can be very misleading, as even for rather high anisotropy energies, the shape and the imaginary part of the susceptibility spectra drastically differ from that of a purely Brownian, or purely Néel relaxing systems.
 
In summary, we can see that the tSW approach can be used effectively to incorporate Néel relaxation and simulate the dynamics associated with that relaxation process for magnetic nanoparticles with $\sigma \geq 5$. The relative importance of the higher-frequency mechanisms not considered in the tSW approach has to be determined on a case-by-case basis. The tSW approach could enable new simulation studies in cases where such mechanisms are not crucial. The price to pay in terms of accuracy is slight, compared to the simulation scale compromises that would be unavoidable with an approach that reproduces also the higher frequency relaxation processes. A major use case of the tSW model would be to simulate systems where $\tau_B$ is high. In this case, a Brownian dynamics approach to thermalize each particle might be impractical. A good example here would be multicore or composite magnetic colloids, where the nanoparticles are embedded in a solid matrix~\cite{eberbeck2012multicore,kuznetsov2022structural,kuznetsov2023multicore}.

\section{Conclusions}\label{sec:concl}
Driven by a long-standing need to incorporate internal magnetization dynamics into coarse-grained computer simulations of magnetic soft matter, we proposed an approach capable of encompassing both Néel and Brownian relaxations in a computationally affordable manner. In this paper, we introduced the Stoner-Wohlfarth thermal model (tSW) and validated it against an ensemble of immobilized, non-interacting, randomly oriented uniaxial particles (solid superparamagnet) and a classical dilute (non-interacting) ferrofluid for various combinations of anisotropy strength and Zeeman energy at a fixed temperature. The choice of systems was dictated by the availability of analytical results, which were used to establish the range of validity of our approach. It was crucial to verify whether and when the tSW model could reproduce both the equilibrium and dynamic properties of magnetic soft matter systems. Our findings indicate that the tSW model is applicable and efficient when the anisotropy of a particle exceeds thermal fluctuations by a factor of five or more. We also applied the tSW model to interacting systems, uncovering and rationalizing a pronounced magnetization decrease in solid superparamagnets due to the dipole-interaction-induced formation of magnetic vortices/holes. The linear dependence of the proportion of these holes on $\xi$ fits most of our data, though its physical meaning and general functional form for arbitrary $\xi$ require further investigation. This work opens possibilities for future studies of magnetic soft matter systems, where the interplay between Néel and Brownian relaxation is often overlooked. We plan to address field-dependent relaxation dynamics within the tSW model in a future study.

\section{Methodology}\label{sec:method}

\subsection{Simulation Method}
We perform MD simulations of a solid superparamagnet and a non-interacting ferrofluid for different anisotropy and Langevin parameter combinations $(\sigma,\xi)$, at a fixed temperature, using the ESPResSo simulation package~\cite{weik2019espresso}. Magnetic nanoparticles are assumed to be mono-disperse, spherical particles with a point dipole, and a characteristic diameter $d$. The solvent was implicitly represented via the Langevin thermostat~\cite{allen2017computer}. The equations of motion are integrated over time $t$ numerically:
\begin{equation}
    \label{eq:lang_trans}
    M_{i} \frac{d \vec{\mbox{$\nu$}}_{i}}{dt} = \vec{\mbox{F}}_{i} - \Gamma_{Tl} \vec{\mbox{$\nu$}}_{i} + 2 \vec{\mbox{$\xi$}}_{i}^{Tl},
\end{equation}
\begin{equation}
    \label{eq:lang_rot}
    I_{i} \frac{d \vec{\mbox{$\omega$}}_{i}}{dt} = \vec{\mbox{$\tau$}}_{i} - \Gamma_{R} \vec{\mbox{$\omega$}}_{i} + 2 \vec{\mbox{$\xi$}}_{i}^{R},
\end{equation}
\noindent where for the $i$-th particle in Eq. \eqref{eq:lang_trans}, $M_{i}$ is, in general, a rank two mass tensor. Since we are simulating isotropic colloids, the mass tensor reduces to a scalar. $\vec{\mbox{F}}_{i}$ is the force acting on the particle; $\vec{\mbox{$\nu$}}_{i}$ denotes the translational velocity. $\Gamma_{Tl}$ denotes the translational friction tensor that once again, due to the isotropy arguments, reduces to a scalar friction coefficient. Finally, $\vec{\mbox{$\xi$}}_{i}^{Tl}$ is a stochastic force modeling the thermal fluctuations of the implicit solvent. Similarly, in Eq. \eqref{eq:lang_rot}, $I_{i}$ denotes $i$-th particle inertia tensor (scalar for a homogeneous sphere), $\vec{\mbox{$\tau$}}_{i}$ is the torque acting on it, $\vec{\mbox{$\omega$}}_{i}$ is the particle rotational velocity. As for the translation, $\Gamma_{R}$ denotes the rotational friction tensor that reduces to a scalar for our colloids, and $\vec{\mbox{$\xi$}}_{i}^{R}$ is a stochastic torque serving for the same purpose as $\vec{\mbox{$\xi$}}_{i}^{Tl}$. Both stochastic terms satisfy the conditions on their time averages~\cite{uhlenbeck1930theory}:
\begin{equation}
\begin{aligned}
    \langle \vec{\mbox{$\xi$}}^{Tl/R} \rangle_t &= 0 \\
    \langle \vec{\mbox{$\xi$}}_{l}^{Tl/R}(t) \vec{\mbox{$\xi$}}_{k}^{Tl/R} (t^\prime) \rangle &= 2\Gamma_{Tl/R} k_B T \delta_{l,k}\delta(t-t^\prime)  ;
\end{aligned} 
\end{equation}
\noindent
where $k,l=x,y,z$.

Forces and torques in Eqs. \eqref{eq:lang_trans} and \eqref{eq:lang_rot} are calculated from interaction potentials. We used periodic boundary conditions, to avoid finite-size effects. Integration of the equations of motion was performed using the velocity Verlet algorithm~\cite{rapaport2004art}, with a timestep of 0.01 (see \ref{ssec::units} for more detail on the simulation units and their relation to experimental values). The tSW calculation, as described in Section \ref{ssec:model}, is done before the force calculation in the velocity Verlet scheme. The initial configuration on our simulations is constructed by randomly placing 500 particles in a cubic simulation box, where the anisotropy axis orientations were uniformly distributed on a surface of a sphere. Simulations length was chosen to be 100$\tau_{char}$, where $\tau_{char}$ denotes the characteristic time of the longest relaxation process we needed to resolve. To obtain statistically significant results, we always present averages over eight independent simulation runs, and over 20 statistically independent snapshots for each simulation. We include the Zeeman energy from the external magnetic field $\vec{H}$:
\begin{equation}
U_H=-\mu_0\sum_{i=0}^N\vec{H}\cdot \vec{\mu}_i.
\label{eq:zee}
\end{equation}

The long-range dipole-dipole pair interactions are simulated using direct summation with two replicas. The performance and reliability of the implementation of tSW is dependent on a robust and highly efficient strategy to find the extrema in Eq.\ref{eq::sw_minim}. For this purpose, we make use of \emph{Nlopt}, a free/open-source library for nonlinear optimization~\cite{NLopt}. Nlopt is a versatile library written in C, with a common interface for a multitude of different optimization strategies. We have specifically settled on using the Method of Moving Asymptotes (MMA) algorithm~\cite{svanberg2002class} to find extrema in Eq.\ref{eq::sw_minim}. MMA is guaranteed to converge to some local minimum and has, throughout our testing, proven to be well suited for optimisation of Eq.\ref{eq::sw_minim}, converging to the correct energy minima very quickly.

\subsection{Reduced units and mapping to physical parameters}\label{ssec::units}

In this subsection, we give an overview of the SI and reduced units used in our simulations. The units are presented in the order we find most instructive. We start with a common reference material choice for the magnetic nanoparticles, namely, magnetite. We assumed that magnetite has uniaxial anisotropy with a magnetic anisotropy constant of $K=10\; kJ/m^3$, saturation magnetization $M_s=480\cdot 10^3\; A/m$ and a core density $\rho_m=5170 \; kg/m^3$. The uniaxial anisotropy assumption is commonly used, but is not strictly correct, as pointed out in~\citet{witt2005three}. In all cases, we assumed that the magnetite core is coated with a 2 \emph{nm} thick oleic acid coating (density $895 kg/m^3$). Given a value of $\sigma$ we want to explore, considering the material values given above, the unit length is chosen to be the diameter of the nanoparticle, $d$, and a unit mass is chosen to be its mass. We choose the particle volume fraction in the simulation box to be $\phi=0.001$. The side length of the simulation box was set to $\approx 64d$, derived based on the chosen $\phi$ and the particle number, $N$. In all cases, we choose the unit energy to be room temperature $T=298.15K$. In other words, we set the reduced temperature of the Langevin thermostat to $k_B T=1$. These three choices also uniquely determine the unit time. From there, we calculate $\tau_D$, where the gyromagnetic ratio is $\mu_0\cdot (1.76\cdot 10^{11}\; s^{-1}T^{-1})$, the damping parameter is chosen to be $\alpha=0.08$ and $\mu_0$ is the vacuum permittivity. We choose $\tau_D/\tau_B=0.01$. Based on this choice, we can derive the density of the implicit fluid and the respective rotational and translational friction coefficients.

\section*{Acknowledgements}

This research has been supported by the Project SAM P 33748. Computer simulations were performed at the Vienna Scientific Cluster (VSC-5). SSK acknowledges the support of Project DEMMON PAT 4120124, and DN MAESTRI (Marie Skłodowska-Curie Actions-Doctoral Networks grant agreement No 101119614). The authors thank Prof. Alexey O. Ivanov for guidance and feedback during the development of this work.
 
\section{Appendix A}

Here, we derive the equilibrium magnetization for an ensemble of immobilized, non-interacting, randomly oriented uniaxial particles. First, let us consider immobilized particles with a predefined orientation of  the anisotropy axes ($\vec{n} = const$).The equilibrium projection of magnetization on the field direction is
\begin{equation}\label{eq::m_h}
m_H = \langle \cos \theta \rangle_{eq} = \int \cos \theta \: W_{eq} \:  \text{d} \vec{e},
\end{equation}
where $\int \ldots d \vec{e}$ is the integral over all possible orientations of the magnetic moment and $W_{eq}$ is the equilibrium (Gibbs/Boltzmann) PDF,
\begin{equation}\label{eq:gibbs_ensembe}
    W_{eq} = \frac{\exp \big( - U / k_B T \big)}{Z(\xi, \sigma, \vec{n})}
\end{equation}
$Z$ is the partition function,
\begin{equation}\label{eq::part_funct}
\begin{aligned}
     Z(\xi, \sigma, \vec{n}) &= \int \exp \big( - U / k_B T \big)  \text{d} \vec{e}\\
     &= \int \exp \big[ \xi \cos \theta + \sigma \left(\vec{e}\cdot\vec{n}\right)^2 \big]  \text{d} \vec{e}. 
\end{aligned}
\end{equation}
Combining Eq.~\ref{eq:gibbs_ensembe} and Eq.~\ref{eq::part_funct}:
\begin{equation}
    m_H= \frac{\partial_{\xi} Z(\xi, \sigma, \vec{n})}{Z(\xi, \sigma, \vec{n})}.
\end{equation}
In practice, it is easier to consider integral Eq.~\ref{eq::part_funct} in a spherical coordinate system, in which $z$-axis coincides with the easy axis
and the field lies perpedicular to the $xz$-plane, i.e., $\vec{n} = (0,0,1)$ and $\vec{H} = H(\sin \psi, 0, \cos \psi)$.  Let us denote magnetic moment angles in this system as $\phi_n$ and $\theta_n$, so $\vec{e} = (\cos \phi_n \sin \theta_n, \sin \phi_n \sin\theta_n, \cos \theta_n)$. Note, in the main text, it was an external field that was aligned along the $z$-axis, and the easy axis was forming an angle $\psi$ with it in the $xz$-plane.
Then Eq.\ref{eq::part_funct} becomes:
\begin{widetext}
\begin{equation}
    \begin{aligned}
Z(\xi, \sigma, \psi) &=
\int^{2 \pi}_0  \text{d} \phi \int^{\pi}_0  \exp \big[\sigma \cos^2 \theta_n + \xi (\cos \theta_n \cos \psi +   \cos \phi_n \sin \theta_n \sin \psi)\big] \sin \theta_n  \text{d} \theta_n \\
&= \int^{\pi}_0  \exp \big(\sigma \cos^2 \theta_n +  \xi \cos \theta_n \cos \psi \big)  \Bigg(\int^{2 \pi}_0 \exp \big[ \xi \cos \phi_n \sin \theta_n \sin \psi\big]   \text{d} \phi_n \Bigg) \sin \theta_n  \text{d} \theta_n \\
&= 2 \pi \int^1_{-1} \exp \big(\sigma x^2\big) \exp \big( \xi  \cos \psi \: x \big) I_0 \Bigg(\xi \sin \psi \sqrt{1 - x^2}\Bigg)  \text{d} x \\
&= 2 \pi \int^1_{-1} \exp \big(\sigma x^2\big)  \Big[ \cosh\big(\xi  \cos \psi \: x\big) + \sinh\big(\xi  \cos \psi \: x\big)  \Big] I_0 \Bigg(\xi \sin \psi \sqrt{1 - x^2}\Bigg)  \text{d} x \\
&= 4 \pi \int^1_{0} \exp \big(\sigma x^2\big)  \cosh\big(\xi  \cos \psi \: x\big)  I_0 \Bigg(\xi \sin \psi \sqrt{1 - x^2}\Bigg)  \text{d} x.
\end{aligned}
\end{equation}
\end{widetext}
where $I_0$ is the modified Bessel function of the first kind of order zero. Analogous expressions for the partition function of an immobilized particle can be found in Refs.~\cite{cregg1999single,elfimova2019static}. 
The derivative is given by:
\begin{widetext}
    \begin{align}
        \partial_{\xi}Z(\xi, \sigma, \psi) = 4 \pi \int^{1}_0 \exp \big(\sigma x^2\big) \Bigg[& x \cos \psi  \sinh\big(\xi  \cos \psi \: x\big)  I_0 \bigg(\xi \sin \psi \sqrt{1 - x^2} \bigg) + \\
        &\sqrt{1 - x^2} \sin \psi \cosh\big(\xi  \cos \psi \: x\big)  I_1 \bigg(\xi \sin \psi \sqrt{1 - x^2} \bigg) \Bigg]  \text{d} x,
    \end{align}
    
\end{widetext}
where $I_1$ is the modified Bessel function of the first kind of order one. To get the magnetization of an ensemble with randomly distributed easy axes, one needs to average $m_H$ over all possible orientations of $\vec{n}$.
\begin{equation}\label{eq:master_solid}
    \begin{aligned}
        m &= \frac{1}{4 \pi} \int m_H(\xi, \sigma, \vec{n})  \text{d} \vec{n} \ = \ \frac{1}{4 \pi} \int^{2\pi}_0  \text{d} \zeta \times\\
        &\times  \int^{\pi}_0 m_H(\xi, \sigma, \psi) \sin \psi  \text{d} \psi = \int^{\pi/2}_0 m_H(\xi, \sigma, \psi) \sin \psi  \text{d} \psi.
    \end{aligned}
\end{equation}
The transition $\int^{\pi}_0  \text{d}\psi \rightarrow 2 \int^{\pi/2}_0  \text{d}\psi$ can be made because the integrand is symmetrical with respect to $\cos \psi$. Also note, that since there is no special direction in the plane orthogonal to the field, the magnetization of a random ensemble must be strictly aligned with the field and the perpendicular component of magnetization must be zero (this is why subscript $H$ can be omitted).
\bibliography{bibliography} 
\end{document}